\documentclass[sigconf]{acmart}
\AtBeginDocument{%
  }

\setcopyright{acmlicensed}
\copyrightyear{2025}
\acmYear{2025}
\acmDOI{XXXXXXX.XXXXXXX}
\acmConference[Conference acronym 'XX]{Make sure to enter the correct
  conference title from your rights confirmation email}{June 03--05,
  2018}{Woodstock, NY}
\acmISBN{978-1-4503-XXXX-X/2018/06}

\usepackage{makecell}
\usepackage{subcaption}
\usepackage{amsthm}
\usepackage{amsmath}

\usepackage{algorithm} 
\usepackage{algpseudocode} 

\algnewcommand{\Let}[2]{\State #1 $\gets$ #2} 
\algrenewcommand{\algorithmiccomment}[1]{\hfill // #1} 
\algnewcommand{\Input}[1]{\State \textbf{Input:} #1}
\algnewcommand{\Output}[1]{\State \textbf{Output:} #1}
\begin{document}

\title{Reinforce Lifelong Interaction Value of User-Author Pairs for Large-Scale Recommendation Systems}
\author{Yisha Li}
\affiliation{%
  \institution{Kuaishou Technology}
  \city{Beijing}
  \country{China}}
\email{liyisha@kuaishou.com}

\author{Lexi Gao}
\affiliation{%
  \institution{Kuaishou Technology}
  \city{Beijing}
  \country{China}}
\email{gaolexi@kuaishou.com}

\author{Jingxin Liu}
\authornote{Corresponding author.}
\affiliation{%
  \institution{Kuaishou Technology}
  \city{Beijing}
  \country{China}}
\email{liujingxin05@kuaishou.com}

\author{Xiang Gao}
\affiliation{%
  \institution{Kuaishou Technology}
  \city{Beijing}
  \country{China}}
\email{gaoxiang12@kuaishou.com}

\author{Xin Li}
\affiliation{%
  \institution{Kuaishou Technology}
  \city{Beijing}
  \country{China}}
\email{lixin05@kuaishou.com}

\author{Haiyang Lu}
\affiliation{%
  \institution{Kuaishou Technology}
  \city{Beijing}
  \country{China}}
\email{luhaiyang@kuaishou.com}

\author{Liyin Hong}
\affiliation{%
  \institution{Kuaishou Technology}
  \city{Beijing}
  \country{China}}
\email{hongliyin@kuaishou.com}

\renewcommand{\shortauthors}{Li et al.}

\begin{abstract}
Recommendation systems (RS) help users find interested content and connect authors with their target audience. 
Most research in RS tends to 
focus either on predicting users' immediate feedback (like click-through rate) accurately 
or improving users' long-term engagement. 
However, they ignore the influence for authors and the lifelong interaction value (LIV) of user-author pairs,
which is particularly crucial for improving the prosperity of social community on different platforms. 
Currently, reinforcement learning (RL) can optimize long-term benefits and has been widely applied in RS. 
In this paper, we introduce RL to \textbf{R}einforce \textbf{L}ifelong \textbf{I}nteraction \textbf{V}alue of \textbf{U}ser-\textbf{A}uthor pairs (RLIV-UA) 
based on each interaction of UA pairs. 
To address the long intervals between UA interactions and the large scale of the UA space, 
we propose a novel Sparse Cross-Request Interaction Markov Decision Process (SCRI-MDP) 
and introduce an Adjacent State Approximation (ASA) method to construct RL training samples. 
Additionally, we introduce Multi-Task Critic Learning (MTCL) to 
capture the progressive nature of UA interactions (click → follow → gift), 
where denser interaction signals are leveraged to compensate for the learning of sparse labels. 
Finally, an auxiliary supervised learning task is designed to enhance the convergence of 
the RLIV-UA model. 
In offline experiments and online A/B tests, 
the RLIV-UA model achieves both higher user satisfaction and higher platform profits 
than compared methods. 
\end{abstract}





\maketitle

\section{Introduction}
\begin{figure}[htbp]
  \centering
  \includegraphics[width=0.75\linewidth]{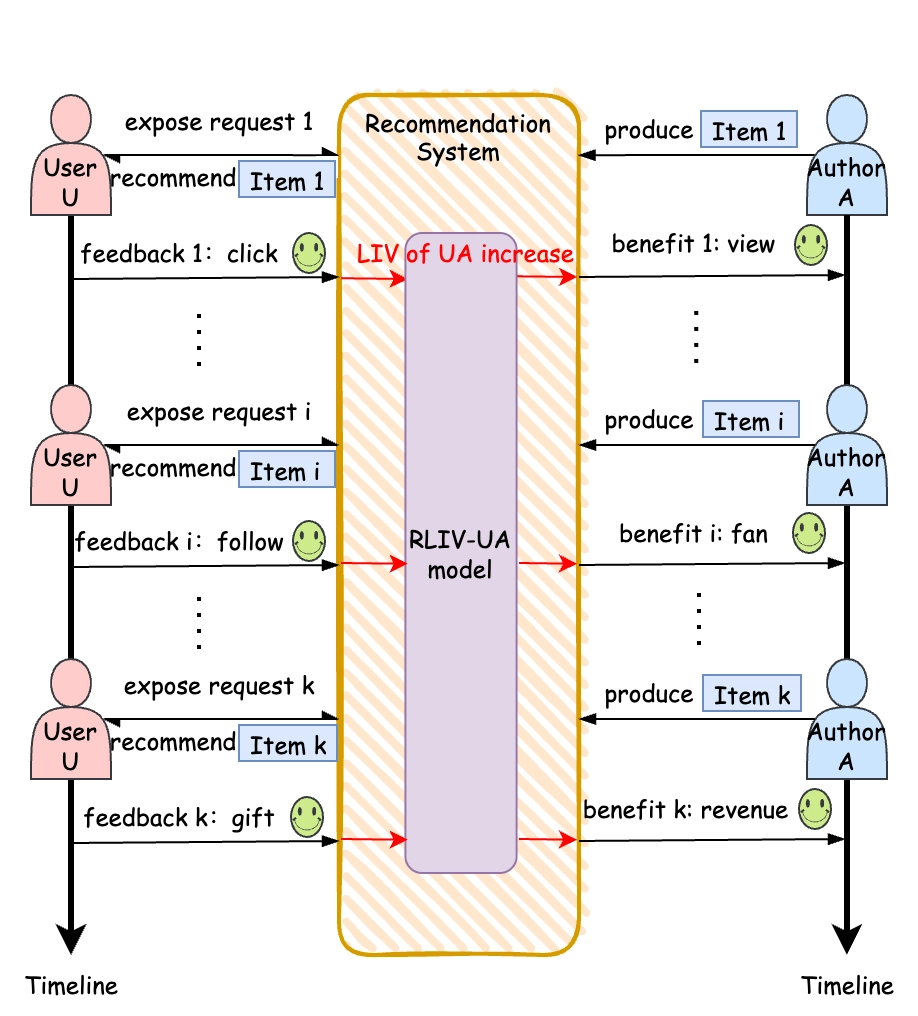}
  \caption{The main process of the proposed RLIV-UA model optimizing the LIV of UA pair based on their interactions.}
  \label{image1}
\end{figure}

The recommendation system (RS) aims to help users discover content aligned with their interests, while simultaneously enabling content authors to reach their target audiences, thereby facilitating fan accumulation and revenue generation \cite{zangerle2022evaluating,jesse2021digital,afsar2022reinforcement}. By fostering repeated, mutually beneficial interactions between users and authors, RS plays a pivotal role in cultivating vibrant platform ecosystems, ultimately driving increased user engagement, traffic, and commercial returns \cite{azaria2013movie, dias2008value, lu2014show, ras2017user, jannach2017price, jannach2019measuring, jannach2020escaping}.

Current research in RS largely falls into two categories. The first focuses on improving the accuracy of immediate user feedback prediction shown in Fig \ref{image1}, such as click-through rate (CTR), at each recommendation request, typically using deep neural networks (DNNs) \cite{linden2003amazon, he2014practical, covington2016deep, zhou2018deep, gu2020hierarchical}. The second category leverages reinforcement learning (RL) to optimize long-term user engagement from \textbf{the user’s perspective} \cite{zhang2022multi, xue2023prefrec, chen2021values, wang2022surrogate, cai2023reinforcing, zheng2018drn, zou2019reinforcement, wang2021learning}, dynamically maximizing session-level or trajectory-level cumulative rewards \cite{sutton2018reinforcement, van2016deep, fujimoto2021minimalist}.

However, both paradigms largely overlook a critical dimension: \textbf{the Lifelong Interaction Value (LIV) of user-author (UA) pairs}. By neglecting the bidirectional, evolving nature of UA relationships, existing methods fail to capture how sustained interactions, from initial discovery to deep loyalty, jointly benefit both parties and, by extension, the platform itself. This omission is particularly consequential, as the long-term stickiness and vitality of platform communities hinge on nurturing these relationships.


As further evidenced in Appendix \ref{appendixB}, we observe a strong positive correlation between the depth and frequency of lifelong UA interactions and key platform-level outcomes, including total revenue and average app usage time. This underscores a fundamental insight: strategically modeling and reinforcing the LIV of UA pairs is not merely a user- or author-centric optimization problem, but a direct pathway to maximizing overall platform prosperity.

To the best of our knowledge, this paper presents the first framework to \textbf{R}einforce \textbf{L}ifelong \textbf{I}nteraction \textbf{V}alue of \textbf{U}ser-\textbf{A}uthor pairs (RLIV-UA) using RL. As shown in Fig. \ref{image1}, RLIV-UA dynamically optimizes the cumulative value of each UA interaction to progressively strengthen the mutual stickiness between user U and author A. Directly applying RL to model LIV, however, introduces significant challenges: (1) UA interactions often span long, irregular time intervals, unlike the dense, request-aligned interactions typically modeled in RS; (2) The combinatorial scale of the UA state space,  driven by hundreds of millions of users and authors, renders it infeasible to store complete historical interaction traces over extended periods (e.g., months).

To address these challenges, we propose a novel \textbf{S}parse \textbf{C}ross-\textbf{R}equest \textbf{I}nteraction \textbf{M}arkov \textbf{D}ecision \textbf{P}rocess (SCRI-MDP) to formally model sparse, long-interval UA dynamics; an \textbf{A}djacent \textbf{S}tate \textbf{A}pproximation (ASA) method to construct practical RL training samples under storage constraints; and a \textbf{M}ulti-\textbf{T}ask \textbf{C}ritic \textbf{L}earning (MTCL) architecture \cite{liu2023multi} that captures the progressive nature of UA relationships (e.g., click → follow → gift), leveraging denser signals to bootstrap learning of sparse, high-value actions. Finally, to combat sample sparsity and label variance, we introduce a supervised learning task to stabilize training and accelerate convergence.

In summary, the key contributions of this work are:
\begin{itemize}
\item We propose RLIV-UA, a novel reinforcement learning framework designed to optimize the lifelong interaction value of user-author pairs in large-scale recommendation systems.
\item We introduce SCRI-MDP, a formalism for modeling sparse, cross-request UA interactions, along with ASA, a practical method for constructing RL training samples under industrial constraints.
\item We design an MTCL architecture to model the hierarchical progression of UA relationships (e.g., from clicks to gifts), augmented by an auxiliary supervised learning task to ensure stable and efficient training.
\item Extensive offline simulations and large-scale online A/B tests demonstrate that RLIV-UA significantly improves both user engagement and author benefits, leading to substantial gains in platform revenue.
\end{itemize}

\section{Problem Formulation}
Existing RL methods in RS often model user behaviors as infinite request-level markov decision process (MDP) \cite{sutton2018reinforcement}. Specifically, the time interval between adjacent states $\Delta=1$ always holds. However, under the UA interaction space, the time interval between the same UA pair's adjacent interactions satisfies $\Delta \geq 1$. For a specific UA pair, the interactions are usually sparse due to the large scale of candidate recommended items. Therefore, we define a novel sparse cross-request interaction MDP to model the LIV of UA pairs:

\begin{definition}[Sparse Cross-Request Interaction MDP]
A \emph{Sparse Cross-Request Interaction MDP} (SCRI-MDP) is a tuple represented as 
$\mathcal{M} = \langle \mathcal{S}, \mathcal{A}, \mathcal{P}, \mathcal{R}, \gamma, \Delta \rangle$, where:
\begin{itemize}
\item \textbf{State space} $\mathcal{S}$: 
Each state $\mathbf{s}^t_{\mathrm{ua}} \in \mathcal{S}$ encodes user features $\mathbf{f}^t_\mathrm{u}$, author features $\mathbf{f}^t_\mathrm{a}$ and interaction features $\mathbf{f}^t_{\mathrm{ua}}$ (e.g., cumulative watch time, gift count).
\item \textbf{Action space} $\mathcal{A}$: 
The action $a^t_{\mathrm{ua}} \in \mathcal{A}$ is the weight of the ranking score at request $t$. 
\begin{equation}
    final\_score = rank\_score * (c + a_{\mathrm{ua}}^t * w)^b
\end{equation}\label{eq1}
where $w,b,c$ are hype-parameters.
\item \textbf{State transition distribution} $\mathcal{P}$: $P(\mathbf{s}^{t+1}_{\mathrm{ua}} \mid \mathbf{s}^t_{\mathrm{ua}}, a^t_{\mathrm{ua}},\Delta^t_{\mathrm{ua}})$ is determined by the time gap $\Delta^t_{\mathrm{ua}} \ge 1$ between consecutive interactions of the same UA pair, skipping intermediate requests for that pair if no interaction occurs.
\item \textbf{Reward function} $\mathcal{R}$: 
To model the progressive changes of UA relationship, several functions are designed to model the long-term reward of different immediate feedback between UA. Let $r_{\mathrm{ua},c},r_{\mathrm{ua},w},r_{\mathrm{ua},f},r_{\mathrm{ua},g}$ be the reward function of click, effective view, follow, and gift labels,  respectively. $C=\{c,w,f,g\}$ denotes the target label set and $n=4$ denotes the cardinality of $C$.
\begin{equation}
    r_{\mathrm{ua},l} = 
    \begin{cases}
1, & \text{the behavior $l$ happens}, l \in C \\
0, & \text{otherwise}
\end{cases}
\end{equation}
\item \textbf{Discount factor} $\gamma \in [0,1)$: Balances immediate and future rewards.
\item \textbf{Interaction gap} $\Delta^t_{\mathrm{ua}}$: Number of global requests between two consecutive interactions of the same UA pair. 
\end{itemize}
\end{definition}

Note that the proposed SCRI-MDP is a specific instantiation of the general semi-Markov Decision Process (semi-MDP) framework \cite{sutton1999between}. In a standard semi-MDP, actions can take variable amounts of time to complete, and state transitions occur upon the completion of these 'macro-actions'. In our case, the 'macro-action' corresponds to the recommendation policy applied between two consecutive interactions of a UA pair. The decision point (state transition) only occurs when an actual UA interaction happens, which naturally fits the semi-MDP paradigm by handling the variable and often long time intervals ($\Delta \geq 1$) between interactions. Furthermore, the proposed SCRI-MDP can be regarded as an augmented MDP to convert  the semi-MDP paradigm into a 'delay-free' standard MDP paradigm. Then we introduces the following theorem.

\begin{theorem}[Equivalence to augmented MDP]
Consider a Sparse Cross-Request Interaction MDP $\mathcal{M} = \langle \mathcal{S}, \mathcal{A}, \mathcal{P}, \mathcal{R}, \gamma, \Delta \rangle$ with interaction gap variable $\Delta^t_{\mathrm{ua}} \geq 1$. Define an augmented state $\bar{s}_t = (s^t_{\mathrm{ua}}, \Delta^t_{\mathrm{ua}})$ where $\Delta^t_{\mathrm{ua}}$ is treated as part of the state vector, forming an augmented state space $\bar{\mathcal{S}} = \mathcal{S} \times \mathbb{N}$. Then $\mathcal{M}$ is equivalent to a standard MDP $\bar{\mathcal{M}} = \langle \bar{\mathcal{S}}, \mathcal{A}, \bar{\mathcal{P}}, \mathcal{R}, \gamma \rangle$, where the transition kernel is defined as:
\begin{equation}
\bar{P}((\mathbf{s}',\Delta') \mid (\mathbf{s},\Delta),a) = P(\mathbf{s}'_{\mathrm{ua}},\Delta'_{\mathrm{ua}} \mid \mathbf{s}_{\mathrm{ua}},\Delta_{\mathrm{ua}},a_{\mathrm{ua}}),
\end{equation}
and the reward function $\mathcal{R}$ remains identical. Any optimal policy $\pi^*$ in $\mathcal{M}$ corresponds to an optimal policy in $\bar{\mathcal{M}}$ under this mapping.
\label{theorem:1}
\end{theorem}
The proof of \ref{theorem:1} is available in the Appendix \ref{prf:1}. Theorem 1 establishes a formal equivalence between the proposed SCRI-MDP and a standard MDP. It provides a rigorous justification for applying conventional MDP-based reinforcement learning algorithms (such as DQN, DDPG, or TD3, which are grounded in standard MDP theory) to solve the SCRI-MDP problem. By demonstrating that the SCRI-MDP can be transformed into an equivalent standard MDP through state augmentation, we ensure that the theoretical convergence properties and optimality guarantees of these well-studied algorithms remain applicable to our framework. This bridges the gap between our novel, problem-specific formulation and the vast body of existing RL theory and practice. 

Based on the aforementioned SCRI-MDP formulation, we introduce a multi-task critic learning (MTCL) architecture to model the LIV of UA pairs by capturing the progressive changes of lifelong UA  relationship. There are $n$ critic networks $Q_{\phi_{k}},k=1,...,n$ with $\phi_k$ as their trainable parameters to optimize the corresponding cumulative rewards $r_{\mathrm{ua},c},r_{\mathrm{ua},w},r_{\mathrm{ua},f},r_{\mathrm{ua},g}$. The final objective function is as follows:
\begin{equation}
 \max_{\phi_{1}, \cdots, \phi_n} \sum_{l \in C} \mathbb{E}\left[\sum_{t=0}^{\infty}{r_{\mathrm{ua},\,l}^t}\right]
\end{equation}
Each critic network is designed to learn from the states of UA pairs and corresponding rewards. Specifically, at one request of user $\mathrm{u}$, RS recommends author $\mathrm{a}$'s item by outputting the action $a^t_\mathrm{ua} \in \mathcal{A}$ which is the $t$-th interaction of $\mathrm{ua}$, then the optimal LIV of $\mathrm{ua}$ pair and action $a^t_\mathrm{ua}$ is defined as:
\begin{equation}
  Q^\ast(\mathbf{s}^t_{\mathrm{ua}}, a^t_{\mathrm{ua}}) = r^t_{\mathrm{ua}} + \gamma \!\!\!\! \sum_{\mathbf{s}^{t+1}_{\mathrm{ua}} \in \mathcal{S}} \!\!\!\! P(\mathbf{s}^{t+1}_{\mathrm{ua}} \mid \mathbf{s}^t_{\mathrm{ua}}, a^t_{\mathrm{ua}}) \!\!\! \max_{a^{t+1}_{\mathrm{ua}}\in \mathcal{A}}\!\!Q^\ast(\mathbf{s}^{t+1}_{\mathrm{ua}}, a^{t+1}_{\mathrm{ua}})
\end{equation}

\section{Methodology}
In this section, we propose the RLIV-UA model to represent the SCRI-MDP in the UA pair state space. Firstly, due to the long time span between adjacent states of UA pair in the SCRI-MDP, we propose the Adjacent State Approximation (ASA) method to \textbf{construct RL training samples}. Then, we introduce the detailed network architecture of multi-task LIV networks and the final online deployment of the RLIV-UA model. The algorithm pseudocode of the RLIV-UA model is presented in Appendix \ref{appdx:algo}.

 

\subsection{Adjacent State Approximation}
At the t-th interaction of a UA pair, the next state $\mathbf{s}^{t+1}_{\mathrm{ua}}$ is delayed until the next interaction which may occur much later. Thus, the traditional RL training sample $(\mathbf{s}^{t}_{\mathrm{ua}}, a^{t}_{\mathrm{ua}}, r^{t}_{\mathrm{ua}}, \mathbf{s}^{t+1}_{\mathrm{ua}})$ cannot be formed in real time due to sparse, infrequent interactions. In industrial-scale recommendation systems, the sheer number of users and authors makes it infeasible to store the state of every UA pair in a key-value database. Instead, systems typically store user behavior as fixed-length sequential logs. Due to finite storage capacity, these sequences are truncated, which often prevents the system from retrieving a UA pair’s prior interaction, especially for infrequent or new relationships. In our online deployment, the success rate of constructing consecutive RL training samples \((\mathbf{s}^{t-1}_{\mathrm{ua}}, a^{t-1}_{\mathrm{ua}}, r^{t-1}_{\mathrm{ua}}, \mathbf{s}^{t}_{\mathrm{ua}})\) by stitching historical logs is only 47\%. Training LIV networks directly on such sparse and fragmented samples would lead to severe model bias, the RL agent learns from an incomplete, non-representative subset of UA trajectories, significantly degrading its ability to optimize long-term, bidirectional value. Moreover, learning from stitched historical tuples inherently delays the model’s ability to act on new relationships. Specifically, the very first interaction between a user and an author cannot be used for training until a \textbf{second} interaction occurs. This creates a critical blind spot: the model is unable to learn how to nurture the \textbf{initiation phase} of a UA relationship.

To address these challenges, we propose the \textbf{A}djacent \textbf{S}tate \textbf{A}pproximation (ASA) mechanism, a parameterized state predictor \(f_\theta\) to reconstruct the missing next state from the current interaction:

\begin{equation}
  \hat{\mathbf{s}}^{t+1}_\mathrm{ua} = f_\theta(\mathbf{s}^t_\mathrm{ua}, a^t_\mathrm{ua}, r^t_\mathrm{ua})
\end{equation}
where the state \(\mathbf{s}^t_{\mathrm{ua}} = \{\mathbf{f}^t_\mathrm{u}, \mathbf{f}^t_\mathrm{a}, \mathbf{f}^t_{\mathrm{ua}}, \Delta^t_{\mathrm{ua}}\}\) is augmented with the interaction gap \(\Delta^t_{\mathrm{ua}}\), and \(\theta\) denotes the trainable parameters of \(f_\theta\).

\begin{figure}[htbp]
  \centering
  \includegraphics[width=1.0\linewidth]{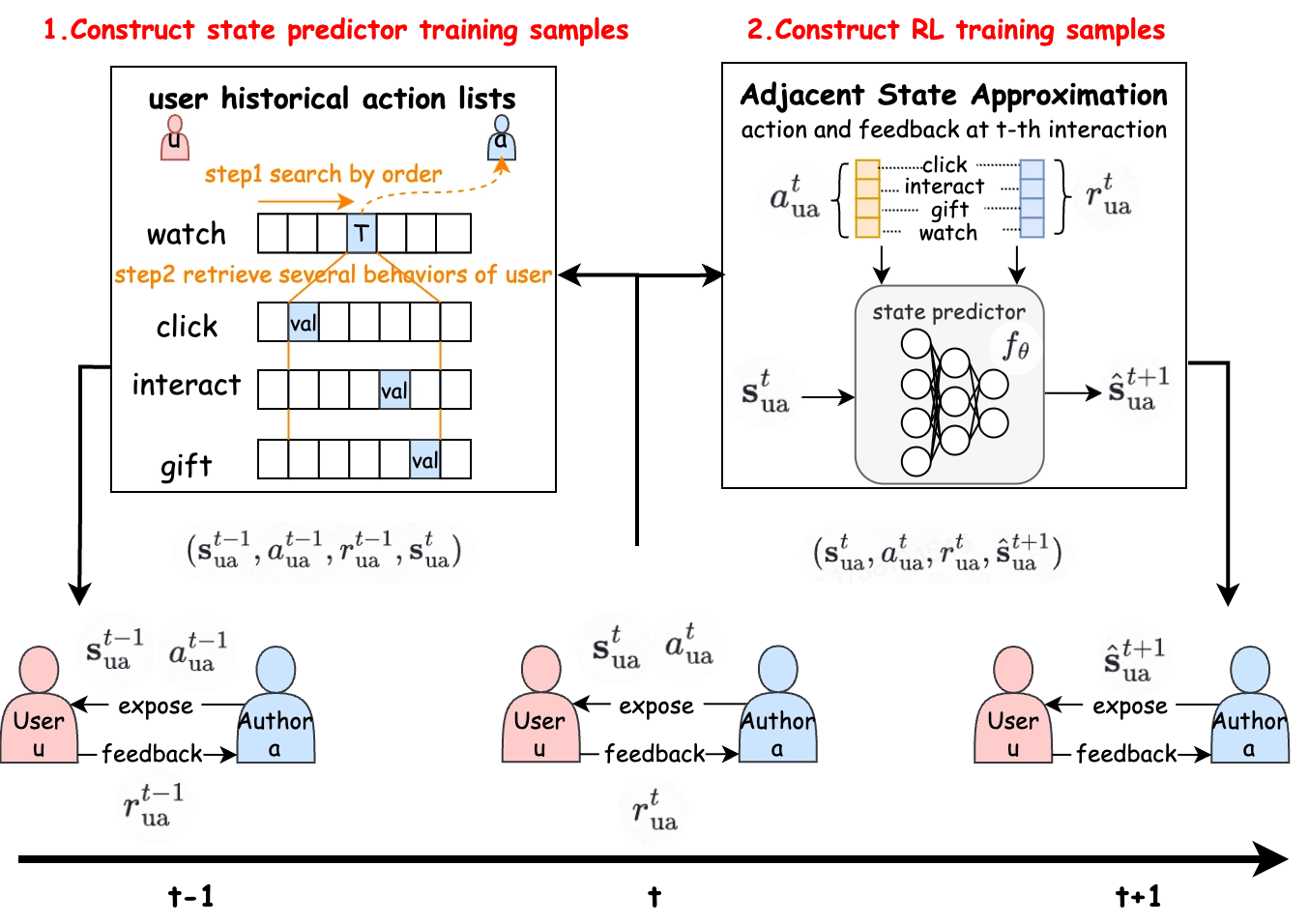}
  \caption{Training sample construction for state predictor and RL model.}
  \label{imagen9}
\end{figure}

As illustrated in Fig. \ref{imagen9}, we first train \(f_\theta\) by retrieving the most recent interaction between the same UA pair from the user’s historical behavior logs. Industrial systems typically store interaction events (e.g., clicks, follows, gifts) as fixed-length sequences. To construct a supervised training sample, we first locate the timestamp \(T\) of the user’s most recent interaction with author \(a\). Since watching an item to completion typically marks the end of an interaction episode, we perform a binary search within a temporal window around \(T\) to retrieve the corresponding action \(a^{t-1}_{\mathrm{ua}}\) and associated feedback signals. This forms the training tuple \((\mathbf{s}^{t-1}_{\mathrm{ua}}, a^{t-1}_{\mathrm{ua}}, r^{t-1}_{\mathrm{ua}}, \mathbf{s}^{t}_{\mathrm{ua}})\), optimized via mean squared error:

\begin{equation}
    \mathcal{L}_{\theta} = \left\| f_\theta(\mathbf{s}^{t-1}_\mathrm{ua}, a^{t-1}_\mathrm{ua}, r^{t-1}_\mathrm{ua}) - \mathbf{s}^{t}_\mathrm{ua} \right\|_2^2
\end{equation}

With ASA, then we can approximate the full RL training sample \((\mathbf{s}^{t}_{\mathrm{ua}}, a^{t}_{\mathrm{ua}}, r^{t}_{\mathrm{ua}}, \hat{\mathbf{s}}^{t+1}_{\mathrm{ua}})\) for \textbf{every} interaction shown in Fig. \ref{imagen9}, not just those with retrievable history. This directly addresses the 53\% stitching failure rate and eliminates the cold-start blind spot for new UA pairs. Crucially, ASA enables the RL agent to learn from \textbf{first-time interactions}, capturing the full lifecycle of UA relationships, from initial discovery to deep, monetizable engagement, and ensuring comprehensive learning across the entire interaction space.

\subsection{Theoretical Analysis}
The proposed adjacent state approximation enables practical training of RL models in industrial recommendation systems. However, a critical question arises: \textit{How does the approximation error in ASA affect the optimality of the learned policy?} To address this, we establish the following theorem that provides an error bound on the Q-function when using approximate state transitions.

\begin{theorem}[Error Bound for Approximate State Transitions]\label{th2}
Let $Q^*$ be the optimal Q-function of the SCRI-MDP $\mathcal{M}$, and let $\hat{Q}$ be the Q-function learned using the ASA with maximum prediction error $\epsilon = \max_{s,a,r} \left|\hat{s}^{t+1} - s^{t+1}\right|$. If the ASA error $\epsilon$ is bounded and the reward function is Lipschitz continuous with constant $L_r$, then the difference between $\hat{Q}$ and $Q^*$ is bounded by:
\begin{equation}
\left|\hat{Q}(s,a) - Q^*(s,a)\right| \leq \frac{L_r \epsilon}{(1-\gamma)^2}
\end{equation}
for all $(s,a) \in \mathcal{S} \times \mathcal{A}$, where $\gamma \in [0,1)$ is the discount factor.
\end{theorem}

The proof of it is available in the Appendix \ref{prf:2}. This theorem provides a theoretical guarantee that the performance degradation caused by ASA is bounded and decreases as the ASA error $\epsilon$ decreases. In practice, we can train the ASA network $f_\theta$ to minimize $\epsilon$, ensuring that the learned policy remains close to optimal. Note that the Lipschitz continuity of the reward function is a standard and widely adopted assumption in theoretical RL literature to ensure the stability and boundedness of value functions. In our implementation, the reward functions are designed as bounded, non-negative functions of user engagement. This design choice inherently promotes smoothness in the reward landscape, making the Lipschitz assumption practically reasonable for our problem setting.

\begin{figure}[htbp]
  \centering
  \includegraphics[width=0.75\linewidth]{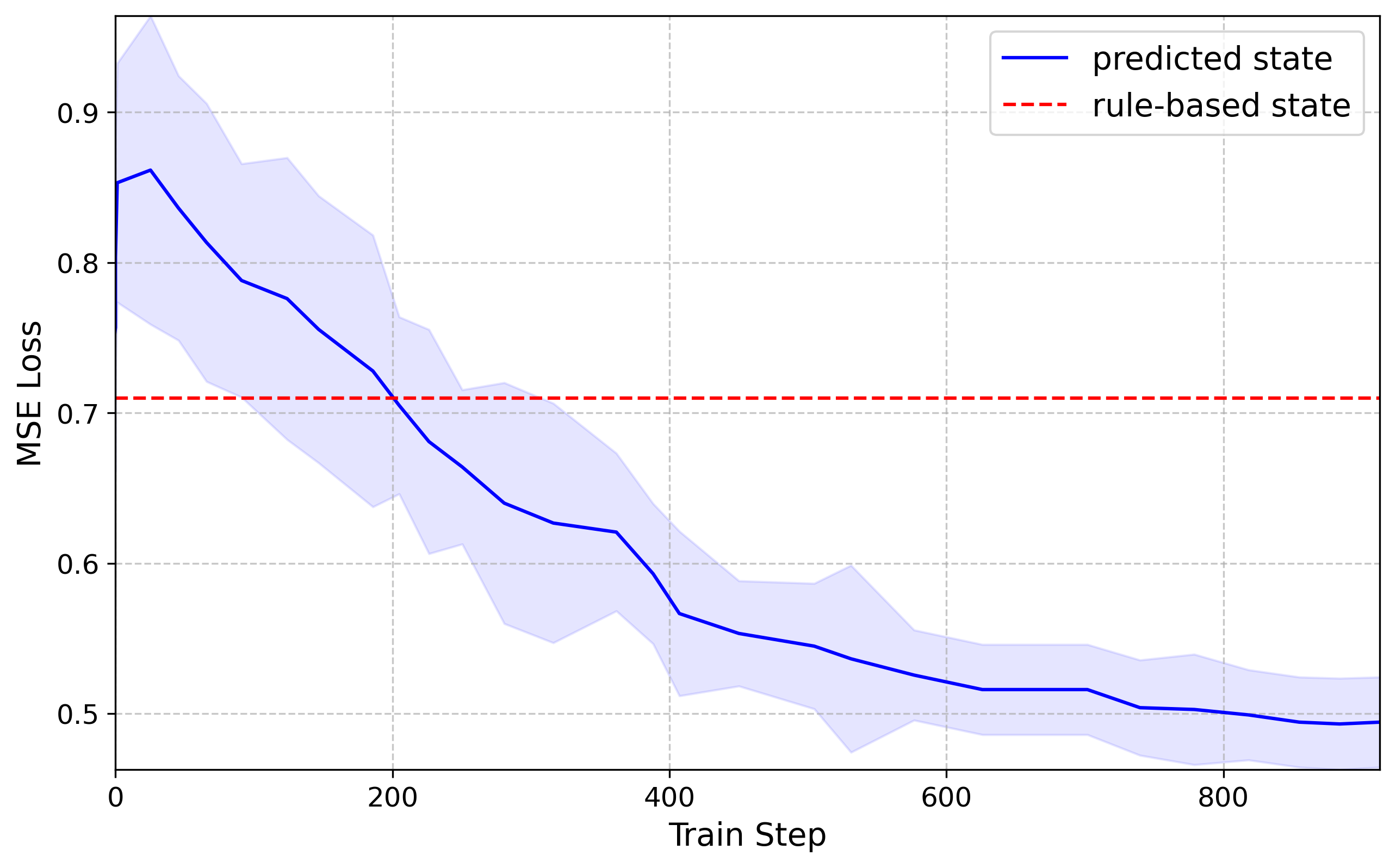}
  \caption{The training process of ASA.}
  \label{img_ASA}
\end{figure}

We further validate the effectiveness of ASA by comparing it against a rule-based baseline, as illustrated in Fig.~\ref{img_ASA}. The ASA method achieves a low MSE loss during training, which directly implies a small approximation error $\epsilon$ in estimating the next state representation $\mathbf{s}^{t+1}$. It further implies that the RL samples generated by ASA are of high fidelity. Consequently, when RLIV-UA is trained on these samples, it provably converges to a near-optimal policy, which satisfies the Q-value performance bound established in Theorem~\ref{th2}. More discussion and offline evaluation results of ASA is detailed in Appendix~\ref{appdx:asa}.

Overall, ASA is a storage-constrained approximation mechanism designed for industrial-scale recommendation systems, where storing complete historical traces for all user-author pairs is infeasible. Unlike model-based RL approaches, e.g. Dreamer \cite{hafner2019dream}, that learn transition dynamics for planning or sample efficiency, ASA serves a singular, pragmatic purpose: to reconstruct valid $(s, a, r, s')$ tuples for RL training when the true next state $s'$ is physically unavailable due to system limitations. Its objective is not to predict environmental dynamics, but to preserve training signal continuity under extreme data sparsity. 

\begin{figure*}[htbp]
  \centering
  \includegraphics[width=0.9\linewidth]{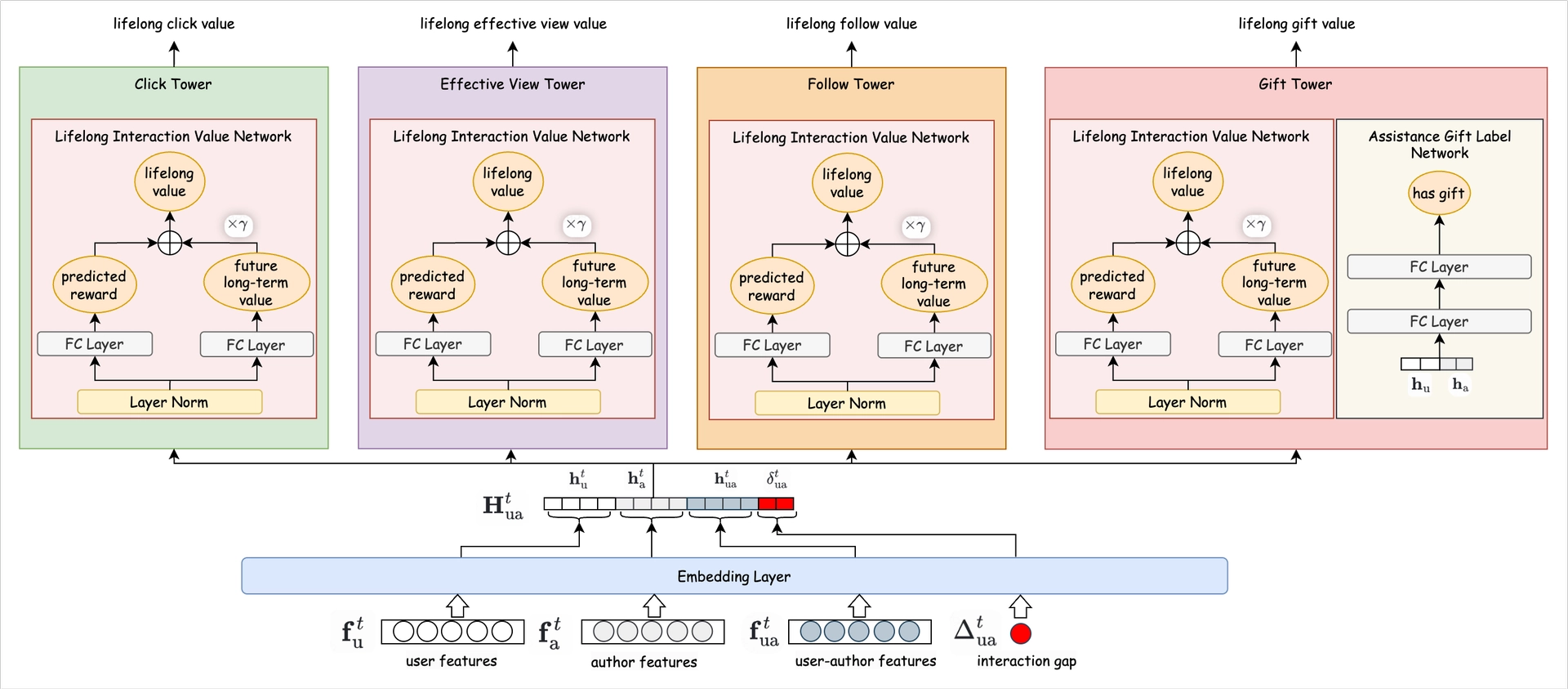}
  \caption{The overall framework of the proposed RLIV-UA model.}
  \label{image3}
\end{figure*}

\subsection{Multi-Task LIV Networks}
 The overall framework of the RLIV-UA model is illustrated in Fig. \ref{image3}. To capture the progressive nature of lifelong user-author relationships, from initial discovery to deep engagement, we design four dedicated LIV networks based on the Multi-Task Critic Learning (MTCL) architecture: Click, Effective View, Follow, and Gift. Crucially, the positive sample rates for these signals vary drastically across tasks: approximately 3\% for Click, 1\% for Effective View, 0.2\% for Follow, and a mere 0.01\% for Gift. This extreme imbalance renders the Follow and Gift labels exceptionally sparse, making direct learning from them highly unstable and inefficient. To address this, MTCL leverages the relatively dense signals (Click and Effective View) as auxiliary tasks to bootstrap and stabilize the learning of the sparse, high-value actions (Follow and Gift), enabling the model to effectively propagate reward signals across the entire interaction hierarchy. For simplicity, we use a single task tower as an example to illustrate the network structure below.

As shown in Fig. \ref{image3}, 
 the current UA state $\left\{\mathbf{f}_\mathrm{u}^{t}, \mathbf{f}_\mathrm{a}^{t}, \mathbf{f}_{\mathrm{ua}}^{t},\Delta^t_{\mathrm{ua}}\right\}$ is fed into a shared embedding layer to obtain corresponding hidden embeddings $\mathbf{h}^{t}_{\mathrm{u}}$, $\mathbf{h}^{t}_{\mathrm{a}}$, $\mathbf{h}^{t}_{\mathrm{ua}}$ and $\delta^t_{\mathrm{ua}}$.
Taking the vector $\mathbf{H}^t_{\mathrm{ua}}=\mathrm{concat}(\mathbf{h}^{t}_{\mathrm{u}},\mathbf{h}^{t}_{\mathrm{a}},\mathbf{h}_{\mathrm{ua}}^t,\delta^t_{\mathrm{ua}})$ as the network input,  $Q_\phi(\mathbf{H}^t_{\mathrm{ua}},a^t_{\mathrm{ua}})$ is denoted as the final LIV of $\mathrm{ua}$ at the $t$-th interaction. To mitigate the problem of value overestimation deviation \cite{van2016deep}, 
double value networks and two corresponding target networks are used to generate the minimum value:
\begin{equation}
  Q_\phi(\mathbf{H}^t_{\mathrm{ua}},a^t_\mathrm{ua})=min(Q_{\phi^1}(\mathbf{H}^t_{\mathrm{ua}},a^t_\mathrm{ua}),Q_{\phi^2}(\mathbf{H}^t_{\mathrm{ua}},a^t_\mathrm{ua}))
\label{eq:5}
\end{equation}

 
Then the corresponding loss function of a LIV network is defined as follows:
\begin{eqnarray}
  \mathcal{L}(\phi)&=&\mathbb{E}_{(\mathbf{s}^t_{\mathrm{ua}}, a^t_{\mathrm{ua}},r^t_{\mathrm{ua}},\hat{\mathbf{s}}^{t+1}_{\mathrm{ua}}) \in D}[(Q_\phi(\mathbf{H}^t_{\mathrm{ua}},a^t_\mathrm{ua})-y)^2] \\
  y &=& r^t_{\mathrm{ua}} + \gamma Q_{\phi'}'(\mathbf{H}^{t+1}_{\mathrm{ua}},\underset{{a^{t+1}_\mathrm{ua}\in \mathcal{A}}}{\mathrm{argmax}}Q_{\phi}(\mathbf{H}^{t+1}_{\mathrm{ua}},a^{t+1}_\mathrm{ua})) \label{eq0}
\end{eqnarray}
where $D$ indicates the sample buffer collected in real time, 
$y$ indicates the target output value of the LIV network, and $Q_{\phi'}'$ represents the output value of the target network with the same structure as the LIV network $Q_\phi$. Note that the network parameter $\phi'$ of $Q_{\phi'}'$ is periodically copied from $\phi$ of $Q_\phi$. 

To reinforce the learning of the sparse gift label of UA pairs in the Gift tower, an assistance MLP network is designed to predict whether user U will gift author A at this interaction. As shown in Fig. \ref{image3}, the user and the author static hidden embeddings $\mathbf{h}_u,\mathbf{h}_a$ are input to the assistance network. The binary cross-entropy loss $\mathcal{L}_A^g$ of the assistance gift binary classification goal is added to the total loss.
\begin{equation}\label{aloss}
    \mathcal{L}_A^g = - \mathbb{E}_{\mathbf{s}^t_{\mathrm{ua}} \in D} \left[ y_g \log(\hat{f}(\mathbf{h}_u,\mathbf{h}_a)) + (1 - y_g ) \log(1 - \hat{f}(\mathbf{h}_u,\mathbf{h}_a)) \right]
\end{equation}
where $y_g$ indicates the true value of whether user U will gift author A at this interaction and $\hat{f}(\mathbf{h}_u,\mathbf{h}_a)$ is the predicted value output by assistance network.

\subsection{Auxiliary Supervised Learning Network}
Previous work \cite{liu2024supervised} finds that the target value $y$ is often dominated by the inaccurate output of the target network $Q_{\phi'}'$ in practice, 
 due to the instability of critic learning in RL. 
 This problem reduces the effectiveness of the real reward $r^t_{\mathrm{ua}}$ in guiding the learning of the value network, since it becomes relatively too small to provide meaningful learning signals. Furthermore, the large scale and extreme sparsity of the UA state space make the RL model even more difficult to converge. 
 
 Therefore, we introduce an auxiliary supervised learning network to regulate the learning of each LIV network $Q_\phi$, preventing a potential divergence of the RL model. 
Specifically, each LIV network is divided into two parts as follows:
\begin{equation}
  Q_\phi(\mathbf{H}^t_{\mathrm{ua}},a^{t}_\mathrm{ua}) := \hat{R}_\eta(\mathbf{H}^t_{\mathrm{ua}},a^{t}_\mathrm{ua})+ \gamma \times \hat{T}_\xi(\mathbf{H}^t_{\mathrm{ua}},a^{t}_\mathrm{ua})
\end{equation}
where $\hat{R}_\eta$ represents the reward prediction network with $\eta$ as its trainable parameters, $\hat{T}_\xi$ is the Q residual network with $\xi$ as its trainable parameters and $\phi=\{\eta,\xi\}$. 

Since the real reward $r^t_{\mathrm{ua}}$ is available based on the $t$-th interaction between $\mathrm{ua}$, 
the predicted reward $\hat{R}_\eta(\mathbf{H}^t_{\mathrm{ua}},a^{t}_\mathrm{ua})$ can be learned by supervised loss. 
Incorporating with the aforementioned clipped double Q-learning \cite{fujimoto2018addressing} shown in Eq. \ref{eq:5}, the auxiliary supervised learning network improves the convergence of the RLIV-UA model.

Hence, the general loss of an LIV network is defined as
\begin{equation}
\mathcal{L}_Q = MSE(\hat{R}_{\eta}(\mathbf{H}^t_{\mathrm{ua}},a^{t}_\mathrm{ua}), r^t_{\mathrm{ua}}) + \sum^{2}_{k=1}MSE(Q_{\phi^k}(\mathbf{H}^t_{\mathrm{ua}},a^{t}_\mathrm{ua}),y) \\
\label{eq:9}
\end{equation}
where the first term denotes the loss for the auxiliary supervised learning network and the second term denotes the original critic learning loss which stops gradients for $\hat{R}_{\eta}$.

Overall, for the whole multi-task LIV networks, the final loss function is defined as follows:
\begin{equation} \label{total_loss_function}
\mathcal{L} = \sum_{l\in C}\mathcal{L}_Q^l + \mathcal{L}_A^g
\end{equation}
where $\mathcal{L}_A^g$ is the assistance gift loss (defined in Eq. \ref{aloss}) and $\mathcal{L}_Q^l$ is the loss function (defined in Eq. \ref{eq:9}), for each label in the target label set $C$, respectively.

\subsection{Online Deployment}
\begin{figure}[h]
  \centering
  \includegraphics[width=0.9\linewidth]{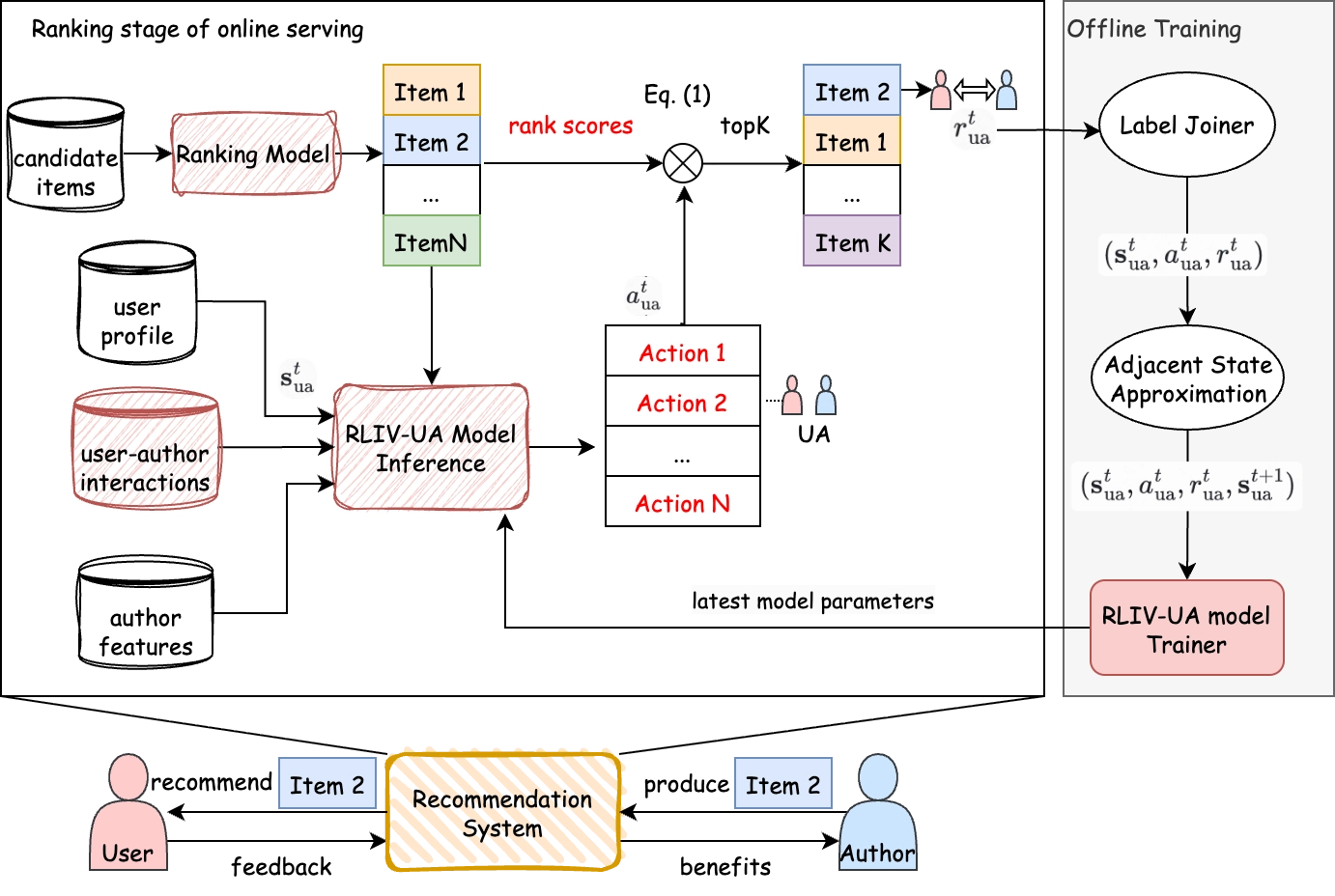}
  \caption{System architecture of the RLIV-UA model in real industrial recommendation scenario.}
  \label{image5}
\end{figure}
In industrial RS, the online system architecture of RLIV-UA model is shown in Fig. \ref{image5}, including the offline training process and the online serving process. In the offline training process, we first utilize the label joiner to merge the UA state features $\mathbf{s}^t_\mathrm{ua}$, the action $a^t_\mathrm{ua}$, and the immediate feedback $r^t_\mathrm{ua}$ at timestamp $t$. Then we leverage the ASA method to approximate the next state $\mathbf{s}^{t+1}_\mathrm{ua}$ and obtain RL training sample $(\mathbf{s}^t_\mathrm{ua}, a^t_\mathrm{ua}, r^t_\mathrm{ua}, \hat{\mathbf{s}}^{t+1}_\mathrm{ua})$. In the online serving process, the RLIV-UA model is deployed as a weight of rank score during the ranking stage of RS to influence the final ordering of candidate items, whose number $N$ is typically less than 300. %

Specifically, at each request of a user, the RLIV-UA model outputs actions with highest LIV value from different task towers for candidate items. For simplicity, there only shows the application process of the action corresponding to a certain tower in Fig \ref{image5}. The candidate items are then ranked by Eq. \ref{eq1} based on their final scores. Finally, the top K candidate items are selected as the final item list and are exposed to the user by order. Then the user interacts with each exposure item to return feedback signals to RS.

In practice, the action $a^t_\mathrm{ua}$ with highest LIV value from different task towers can be selectively applied based on the actual optimization goal, such as improving long-term user retention or maximizing platform revenue. To explore different actions for ranking, the $\epsilon$-greedy strategy is applied online.


\section{Experiments}
The RLIV-UA model is evaluated in an offline simulated recommendation scenario to compare its performance with state-of-the-art models and RLIV-UA variants. It is also applied in two real industrial recommendation scenarios to verify its effectiveness in large-scale industrial RS through online A/B tests. 

\begin{table*}[htbp]
  \centering
  \caption{Overall performance of all compared models in offline recommendation scenario.}
  \label{tab:offlinePerformance}
  \fontsize{8}{11}\selectfont
  \begin{tabular}{lllll}
    \hline
    \textbf{Models} & \textbf{Session Length} & \textbf{Watch Time} & \textbf{CTR} & \textbf{Diversity}   \\
    \hline
    RankingModel & 2.0132& 59.4812 & \textbf{0.5948}  & 0.0629   \\
    CQL & 2.2660& 58.0753 & 0.5125& 0.1727  \\
    DQN & 5.3744& 125.8436& 0.4074& 0.5801  \\
    TD3 & 4.2519& 79.0432& 0.4258& 0.4929  \\
    TD3-UA & 4.9651& 102.0916& 0.4465& 0.7161  \\
    FeedRec & 6.7420& 118.8468& 0.4363& 0.7059  \\
    RLUR & 6.6810& 151.0550& 0.4519& 0.7206  \\
    \hline
    RLIV-UA(w/o MT \& SL) & 7.2940& 188.0499 & 0.5229& 0.7994  \\
    RLIV-UA(w/o MT) & 9.2800& 248.1804& \textbf{0.5340}& 0.8746  \\
    RLIV-UA & \textbf{12.8860}& \textbf{377.4867}& 0.5015& \textbf{0.8827}  \\
    \hline
    \end{tabular}
\end{table*}

\subsection{Experimental Setup}
\subsubsection{Dataset and Evaluation Metrics}
We adopt the KuaiRand dataset \cite{gao2022kuairand} as the foundation for our offline experiments. This public dataset, collected from the Kuaishou app, contains interactions from 27,285 users across 32,038,725 items, resulting in hundreds of millions of UA interaction records. We train an offline user simulator on this dataset to serve as a controllable environment that mimics real user behavior. Specifically, upon receiving a recommended item, the simulator generates immediate feedback signals, such as clicks, watch duration, and comments, and subsequently decides whether to issue the next request based on a probabilistic quitting mechanism, similar to the approach described in \cite{zhang2024unex}.

To comprehensively evaluate model performance, we assess three key dimensions: \textbf{user satisfaction}, \textbf{author benefits}, and \textbf{platform profitability}. The detailed illustration of all evaluation metrics is available in Appendix \ref{appdx:metrics}.

\begin{itemize}
\item \textbf{User Satisfaction} is measured by: Session Length (number of requests per session), Watch Time (total viewing duration per session), and CTR (click-through rate).
\item \textbf{Author Benefits} are quantified via: Diversity (variety of recommended authors/content) and New Fans (number of new followers acquired by authors).
\item \textbf{Platform Profits} are evaluated using: UA Count (number of UA pairs exhibiting deep relationships), Weekly Gifted Users (number of users sending virtual gifts per week), Total Revenue, App Usage Time, and Weekly Retention (proportion of users returning weekly).
\end{itemize}

\subsubsection{Compared Methods}
We compare our method against a suite of representative baselines, including the classic \textbf{RankingModel} and four established reinforcement learning approaches: \textbf{CQL} \cite{kumar2020conservative}, \textbf{DQN} \cite{van2016deep}, \textbf{TD3} \cite{fujimoto2021minimalist}, \textbf{TD3-UA} (Using TD3 to model the SCRI-MDP and the action is a continuous weight value), \textbf{FeedRec} \cite{zou2019reinforcement} and \textbf{RLUR} \cite{cai2023reinforcing}. To further validate the contribution of each component in RLIV-UA, we also evaluate four ablated variants: \textbf{RLIV-UA(w/o MT)} (without Multi-Task learning), \textbf{RLIV-UA(w/o MT \& SL)} (without Multi-Task and Supervised Learning), \textbf{RLIV-UA(w/o AL \& SL)} (without Auxiliary Gift Label Network and Supervised Learning), and \textbf{RLIV-UA(w/o SL)} (without Supervised Learning). The implementation details is available in Appendix \ref{appendix:impl}.

Note that the assistance gift label network is not applied in offline experiments, because there is no gift signals in KuaiRand dataset.
The platform profits metrics are only used in online A/B experiments. All models are trained to convergence, and their results are the averaged performance of the last 10 epochs. Moreover, we use the follow LIV and gift LIV in Kwai and use the watch time LIV in Kuaishou and offline experiments.

\subsection{Performance Comparison}

The overall performance of different models in offline experiment is shown in Table \ref{tab:offlinePerformance}. 
The traditional RankingModel achieves the best performance in CTR since it can predict which item has the greatest probability to be clicked. 
However, it is not suitable for improving the long-term user engagement such as session length and watch time.
The offline model CQL learned from the historical samples can achieve some diversity. 
Compared with traditional RankingModel and the offline RL model CQL,
 most RL-based models achieve better performance in long-term metric session length at the expense of immediate feedback like low CTR, resulting in similar watch time. 
By adding another Q network learning from heuristic rewards, the RLUR model can improve all the long-term metrics including session length and watch time. 
 The proposed RLIV-UA model achieves the best performance in session length and watch time and achieves the third high value in CTR, 
 which reflects that the RLIV-UA model can balance immediate feedback and long-term feedback to improve long-term user engagement by modeling the LIV of UA pairs. Moreover, the RLIV-UA model achieves the best performance in diversity which reflects that modeling the LIV of UA pairs 
can more accurately recommend items of different authors to target users, 
rather than blindly recommend different items. 

\begin{figure}[htbp]
  \centering
  \includegraphics[width=0.75\linewidth]{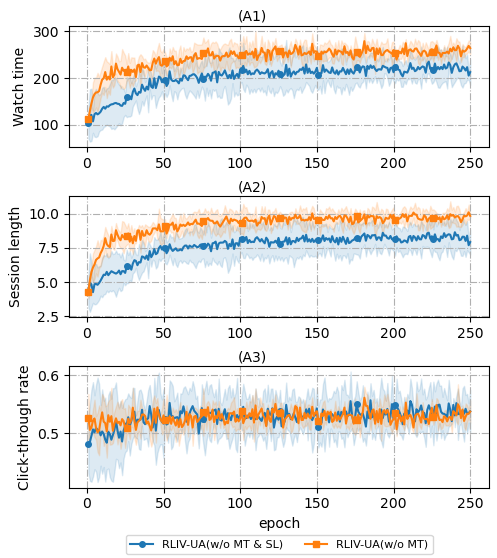}
  \caption{The learning process of RLIV-UA(w/o MT \& SL) and RLIV-UA(w/o MT) over 10 rounds of training where the shaded areas correspond to the standard deviations.}
  \label{image66}
\end{figure}

\subsection{Ablation Study}

The overall performance of RLIV-UA variations in offline recommendation scenario are shown in Table \ref{tab:offlinePerformance}. 

Firstly, compared with the RLIV-UA(w/o MT \& SL) variation, the RLIV-UA(w/o MT) achieves relatively high improvement in session length, watch time and diversity, which reflects that
 the  auxiliary supervised learning task can help model learn the LIV of UA pairs more accurately. 
 
Furthermore, the RLIV-UA achieves the best performance under both session length, watch time and diversity metrics, which indicates the effectiveness of the multi-task critic learning architecture. 
 
 As shown in Fig. \ref{image66}, with the auxiliary supervised learning task, the variance of RLIV-UA(w/o MT) is much lower than that of RLIV-UA(w/o MT \& SL) under watch time, session length and CTR metrics. It demonstrates the learning process of RLIV-UA(w/o MT) model is more stable, and the auxiliary supervised learning task is effective for enhancing the stability of model training.

\begin{figure}[htbp]
  \centering 
  \begin{subfigure}[t]{0.23\textwidth}
    \centering
    \includegraphics[width=\textwidth]{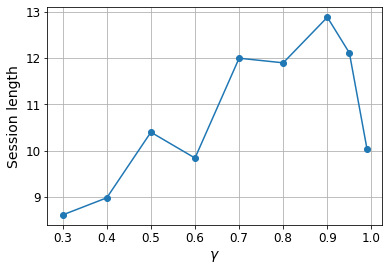}
     \caption{The session length of RLIV-UA with different $\gamma$.}
    \label{fig:sub1}
  \end{subfigure}
  \hfill
  \begin{subfigure}[t]{0.23\textwidth}
    \centering
    \includegraphics[width=\textwidth]{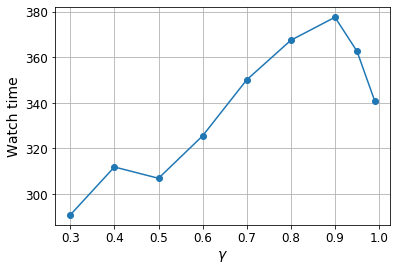}
     \caption{The watch time of RLIV-UA with different $\gamma$.}
    \label{fig:sub2}
  \end{subfigure}
  \caption{The parameter sensitivity experiment of $\gamma$.}
  \label{offlineg}
\end{figure}

As shown in Figure \ref{offlineg}, the RLIV-UA model with $\gamma=0.9$ achieves the best performance in session length 
and watch time and all the models achieve better performance compared with baselines, which shows the RLIV-UA model is not sensitive to the parameter $\gamma$.

\begin{table*}[htbp]
  \centering
  \caption{Results of the revenue experiment on Kwai live-stream feed.}
  \label{tab:onlinerevenue}
  \fontsize{8}{11}\selectfont
  \begin{tabular}{lllll}
    \hline
    \textbf{Models} & 
    \makecell[l]{\textbf{UA Count}}
    & 
    \makecell[l]{\textbf{Weekly Gifted Users}}
    & 
    \makecell[l]{\textbf{Total Revenue}}
    & 
    \makecell[l]{\textbf{New Fans}} \\
    \hline
    RankingModel & - & - & - & -  \\
    RLIV-UA(w/o AL \& SL) & +2.00\% CI:[0.97\%, 5.08\%]& +2.25\% CI:[0.46\%, 4.08\%]& +11.90\% CI:[5.48\%, 32.32\%]& +2.74\% CI:[0.60\%, 4.92\%]  \\
    RLIV-UA(w/o SL) & +3.58\% CI:[1.18\%, 5.36\%]
& +3.42\% CI:[0.23\%, 5.11\%]& +23.81\% CI:[15.93\%, 37.52\%] & +6.10\% CI:[0.10\%, 10.15\%]\\
    RLIV-UA(w/o ASA) & +2.02\% CI:[1.71\%, 5.74\%]
& +1.68\% CI:[0.26\%, 3.10\%]
& +15.21\% CI:[0.36\%, 21.34\%]& +5.22\% CI:[0.36\%, 10.08\%]
  \\
    RLIV-UA & \textbf{+5.85\% CI:[1.34\%, 8.93\%]}& \textbf{+5.37\% CI:[0.11\%, 8.71\%]}& \textbf{+38.64\% CI:[13.43\%, 60.31\%]}& \textbf{+8.28 CI:[0.91\%, 14.38\%]}  \\
    \hline
    \end{tabular}
\end{table*}

\subsection{Online A/B Experiments}
The proposed RLIV-UA model and compared methods were deployed on the Kwai live-streaming feed, a platform serving over 100 million users and 1 million content authors, to optimize platform revenue from July to October 2024. For rigorous evaluation, we randomly assigned 20\% of users (over 20 million) to the treatment group, with the remaining 80\% serving as the control. All reported confidence intervals (CIs) in Table \ref{tab:onlinerevenue} are 95\% two-sided CIs, computed based on the Central Limit Theorem, ensuring statistical robustness given the massive sample size.

As shown in Table \ref{tab:onlinerevenue}, the performance of RLIV-UA and its ablated variants improves progressively: RLIV-UA(w/o AL\&SL) → RLIV-UA(w/o SL) → full RLIV-UA. This consistent uplift across key business metrics, including Total Revenue +38.64\%, Weekly Gifted Users +5.37\%, UA Count +5.85\%, and New Fans +8.28\%, demonstrates the practical effectiveness and additive value of each component in our framework. Notably, we also evaluate a variant, RLIV-UA(w/o ASA), which forgoes the ASA module and instead constructs RL training samples using only stitched historical interactions. This variant underperforms the full RLIV-UA model across all metrics, highlighting ASA’s critical role in mitigating sample sparsity and bias. Without ASA, the model cannot learn from first-time or infrequent UA interactions, severely limiting its ability to optimize lifelong value. This confirms ASA is not optional, but essential for industrial-scale deployment.

To further validate generalizability, we deployed RLIV-UA on the Kuaishou short-video feed from November to December 2024, again using a 20\% user cohort for treatment. Results show statistically significant improvements in long-term user engagement: +0.131\% in average watch time, +0.083\% in daily app usage duration, and +0.021\% in weekly retention rate. Notably, in a system of Kuaishou’s scale, even a 0.02\% gain in retention is considered highly significant, reflecting millions of additional active users. These results confirm that RLIV-UA not only drives immediate revenue growth but also fosters sustainable, long-term platform health by strengthening user-author relationships.

\section{Related work}
\subsection{RL-Based Recommendation Systems}
\cite{sunehag2015deep} is the earliest work 
that tries to alternate multitask learning ranking model with RL model 
using DQN to learn the value of all items in the recommended list. 
Similarly, Chen et al. \cite{chen2019top} employ a policy-gradient approach in RS and 
Zhao et al. \cite{zhao2018deep} develop an actor-critic approach for recommending a page of items. 
However, they are not applied in a real-world recommendation environment with large amount of users and items. 
Then, more research \cite{gauci2018horizon, zheng2018drn} aims to apply the RL model in reality as a substitute with simple network structure. 
In order to handle the huge number of candidate items, 
SlateQ \cite{ie2019slateq} is proposed to decompose the value of item list into the sum of value of each item under some assumptions. 
Other literatures \cite{deffayet2023generative,liu2023exploration} use contrastive learning to overcome 
the curse of dimensionality whose model structures are more complex.

\subsection{Long-Term User Engagement in Recommendation Systems}
In order to consider the long-term user engagement rather than user's immediate feedback, 
some research has increasingly focused on the sequential patterns of user behavior by employing temporal models, 
such as hidden Markov models and recurrent neural networks \cite{rendle2010factorizing,campos2014time,he2016fusing,sahoo2012hidden,tan2016improved,wu2017recurrent}.
Besides, some research \cite{gauci2018horizon,shani2005mdp,taghipour2007usage} use RL to make a long-term planning. 
However, all the methods are too complex to be applied in practice. 
\cite{zou2019reinforcement} propose a hierarchical LSTM based Q network to model the complex user behavior 
and design an S-network to simulate the environment avoiding the instability. 
\cite{chen2021values} inspired by exploration research \cite{mnih2016asynchronous,bellemare2016unifying} in RL use a series of exploration methods to improve user experience.
\cite{wang2022surrogate} carefully design the reward function through data analysis to connect the long-term rewards with immediate feedback. 
While \cite{xue2023prefrec} propose a framework for learning preferences from user historical behavior sequences, 
specifically using preferences to automatically train a reward function in an end-to-end manner.
Considering all the above methods' action is to select an item list which may be not practical when the number of item and user is large, 
\cite{cai2023reinforcing} aim to optimize the weights of each predicted user feedback  when ranking items under the long-term rewards 
with designed heuristic rewards to overcome the latency and sparsity of long-term rewards.

\section{Conclusion}
In this paper, we propose a novel lifelong interaction value model for user-author pairs, i.e. RLIV-UA, based on RL. 
Firstly, the interactions of UA pairs via RS is modeled as a sparse cross-request interaction markov decision process. To solve the long time interval and large scale of UA's interactons, an adjacent state approximation method is designed to build the RL training sample. 
Besides, to capture the progressive changes of lifelong UA relationship, a multi-task critic learning architecture is employed to utilize denser interaction signals to compensate for sparse labels. 
Moreover, an auxiliary supervised learning task is designed to improve the convergence of the RLIV-UA model in large-scale RS.
Finally, in both offline environments and online A/B tests, the experiment results show that 
the proposed RLIV-UA model performs better under both user satisfaction metrics and author benefits metrics, resulting in higher platform profits, 
compared with other models. 

As future work, we plan to explore the application of RLIV-UA’s core components, particularly the SCRI-MDP formulation, ASA method, and MTCL architecture , in other domains such as e-commerce (user-merchant) and music streaming (user-artist), where optimizing sparse, long-term, bidirectional value is equally critical. This will require careful adaptation of reward functions and interaction definitions to fit each domain’s unique dynamics.

\bibliographystyle{ACM-Reference-Format}
\bibliography{sample-base}

\appendix
\section{Proofs}
\subsection{Equivalence to augmented MDP}
\label{prf:1}
\begin{theorem}[Equivalence to augmented MDP]
Consider a Sparse Cross-Request Interaction MDP $\mathcal{M} = \langle \mathcal{S}, \mathcal{A}, \mathcal{P}, \mathcal{R}, \gamma, \Delta \rangle$ with interaction gap variable $\Delta^t_{\mathrm{ua}} \geq 1$. Define an augmented state $\bar{s}_t = (s^t_{\mathrm{ua}}, \Delta^t_{\mathrm{ua}})$ where $\Delta^t_{\mathrm{ua}}$ is treated as part of the state vector, forming an augmented state space $\bar{\mathcal{S}} = \mathcal{S} \times \mathbb{N}$. Then $\mathcal{M}$ is equivalent to a standard MDP $\bar{\mathcal{M}} = \langle \bar{\mathcal{S}}, \mathcal{A}, \bar{\mathcal{P}}, \mathcal{R}, \gamma \rangle$, where the transition kernel is defined as:
\begin{equation}
\bar{P}((\mathbf{s}',\Delta') \mid (\mathbf{s},\Delta),a) = P(\mathbf{s}'_{\mathrm{ua}},\Delta'_{\mathrm{ua}} \mid \mathbf{s}_{\mathrm{ua}},\Delta_{\mathrm{ua}},a_{\mathrm{ua}}),
\end{equation}
and the reward function $\mathcal{R}$ remains identical. Any optimal policy $\pi^*$ in $\mathcal{M}$ corresponds to an optimal policy in $\bar{\mathcal{M}}$ under this mapping.
\end{theorem}

\begin{proof}
We aim to prove that an optimal policy in the SCRI-MDP $\mathcal{M}$ corresponds to an optimal policy in the augmented MDP $\bar{\mathcal{M}}$. We proceed via \textbf{proof by contradiction}.

Let $\pi^* : \mathcal{S} \times \mathbb{N} \to \mathcal{A}$ be an optimal deterministic policy for the SCRI-MDP $\mathcal{M}$. By definition, for all $(s, \Delta) \in \mathcal{S} \times \mathbb{N}$ and for any other policy $\pi$, the value function satisfies:
\begin{equation}
V^{\pi^*}(s, \Delta) \geq V^{\pi}(s, \Delta) 
\end{equation}

Now, consider the corresponding policy $\bar{\pi}^*$ in the augmented MDP $\bar{\mathcal{M}}$, defined such that $\bar{\pi}^*(a |\bar{s}) = \pi^*(a | s, \Delta)$ for $\bar{s} = (s, \Delta)$. Assume, for the sake of contradiction, that $\bar{\pi}^*$ is \textbf{not} optimal in $\bar{\mathcal{M}}$. This implies that there exists another policy $\bar{\pi}' : \bar{\mathcal{S}} \to \mathcal{A}$ and some augmented state $\bar{s}_0 = (s_0, \Delta_0)$ such that:
\begin{equation}
V^{\bar{\pi}'}(s_0, \Delta_0) > V^{\bar{\pi}^*}(s_0, \Delta_0)
\end{equation}

Define a policy $\pi'$ for the original SCRI-MDP $\mathcal{M}$ by $\pi'(a | s, \Delta) := \bar{\pi}'(a \mid (s, \Delta))$. Since both MDPs share the same reward function $\mathcal{R}$, discount factor $\gamma$, and the dynamics of $\mathcal{M}$ are fully encoded in $\bar{\mathcal{P}}$ through the inclusion of $\Delta$ in the state, the expected return from any initial state-action pair is identical under corresponding policies. Specifically, the Bellman equations for both frameworks satisfy:

For the SCRI-MDP:
\begin{equation}
\begin{aligned}
Q^{\pi}(s, \Delta, a) &= \mathcal{R}(s, \Delta, a) +\\
&\gamma \sum_{s',\Delta'} P(s', \Delta' \mid s, \Delta, a)\sum_{a'}\pi(a'\mid s',\Delta') Q^{\pi}(s', \Delta', \pi)
\end{aligned}
\end{equation}

For the augmented MDP:
\begin{equation}
\begin{aligned}
&Q^{\bar{\pi}}((s, \Delta), a) = \mathcal{R}(s, \Delta, a) + \\
& \gamma \sum_{s',\Delta'} \bar{P}((s', \Delta') \mid (s, \Delta), a) \sum_{a'}\bar{\pi}(a'\mid s',\Delta')Q^{\bar{\pi}}((s', \Delta'), \bar{\pi}))
\end{aligned}
\end{equation}

Since $\bar{P}((s', \Delta') \mid (s, \Delta), a) = P(s', \Delta' \mid s, \Delta, a)$ by construction, and the reward function is identical, it follows that:

\begin{equation}
V^{\pi'}(s, \Delta) = V^{\bar{\pi}'}(s, \Delta) \quad \text{and} \quad V^{\pi^*}(s, \Delta) = V^{\bar{\pi}^*}(s, \Delta)    
\end{equation}
for all $(s, \Delta) \in \mathcal{S} \times \mathbb{N}$.

Substituting these identities into our earlier inequality yields:

\begin{equation}
V^{\pi'}(s_0, \Delta_0) > V^{\pi^*}(s_0, \Delta_0)
\end{equation}

This contradicts the assumption that $\pi^*$ is optimal for $\mathcal{M}$. Therefore, the assumption that $\bar{\pi}^*$ is not optimal in $\bar{\mathcal{M}}$ must be false. Hence, $\bar{\pi}^*$ is indeed an optimal policy for $\bar{\mathcal{M}}$.

Conversely, let $\bar{\pi}^*$ be an optimal deterministic policy for the augmented MDP $\bar{\mathcal{M}}$. Then, we have:

\begin{equation}
V^{\bar{\pi}^*}(\bar{s}) \geq V^{\bar{\pi}}(\bar{s})
\end{equation}
for all $\bar{s} \in \bar{\mathcal{S}}$ and for any other policy $\bar{\pi}$.

Define a policy $\pi^*$ for the SCRI-MDP $\mathcal{M}$ by $\pi^*(a | s, \Delta) := \bar{\pi}^*(a \mid (s, \Delta))$. Assume, for contradiction, that $\pi^*$ is not optimal in $\mathcal{M}$. Then there exists another policy $\pi'$ and some state $(s_0, \Delta_0)$ such that:
\begin{equation}
V^{\pi'}(s_0, \Delta_0) > V^{\pi^*}(s_0, \Delta_0)
\end{equation}

Define $\bar{\pi}'$ for $\bar{\mathcal{M}}$ by $\bar{\pi}'(a \mid (s, \Delta)) := \pi'(a \mid s, \Delta)$. Using the same argument as before, we have:
\begin{equation}
V^{\bar{\pi}'}(s_0, \Delta_0) > V^{\bar{\pi}^*}(s_0, \Delta_0)
\end{equation}
which contradicts the optimality of $\bar{\pi}^*$ in $\bar{\mathcal{M}}$. Therefore, $\pi^*$ must be optimal in $\mathcal{M}$.

This establishes a one-to-one correspondence between optimal policies in $\mathcal{M}$ and $\bar{\mathcal{M}}$. Consequently, the SCRI-MDP $\mathcal{M}$ is equivalent to the standard MDP $\bar{\mathcal{M}}$ under the augmented state formulation, completing the proof.
\end{proof}

\subsection{Error Bound for ASA}
\label{prf:2}
\begin{theorem}[Error Bound for Approximate State Transitions]
Let $Q^*$ be the optimal Q-function of the SCRI-MDP $\mathcal{M}$, and let $\hat{Q}$ be the Q-function learned using the ASA with maximum prediction error $\epsilon = \max_{s,a,r} \left|\hat{s}^{t+1} - s^{t+1}\right|$. If the ASA error $\epsilon$ is bounded and the reward function is Lipschitz continuous with constant $L_r$, then the difference between $\hat{Q}$ and $Q^*$ is bounded by:
\begin{equation}
\left|\hat{Q}(s,a) - Q^*(s,a)\right| \leq \frac{L_r \epsilon}{(1-\gamma)^2}
\end{equation}
for all $(s,a) \in \mathcal{S} \times \mathcal{A}$, where $\gamma \in [0,1)$ is the discount factor.
\end{theorem}

\begin{proof}
Let's denote the Bellman operator for the true MDP as $\mathcal{T}$, and for the approximate MDP with ASA as $\hat{\mathcal{T}}$. The Bellman operators are defined as:

\begin{equation}
\begin{aligned}
\mathcal{T}Q(s,a) &= r(s,a) + \gamma \mathbb{E}_{s' \sim P(\cdot|s,a)}\left[\max_{a'} Q(s',a')\right] \\
\hat{\mathcal{T}}Q(s,a) &= r(s,a) + \gamma \mathbb{E}_{\hat{s}' \sim \hat{P}(\cdot|s,a)}\left[\max_{a'} Q(\hat{s}',a')\right]
\end{aligned}
\end{equation}
where $P$ is the true transition probability and $\hat{P}$ is the approximate transition probability induced by ASA.

The fixed points of these operators are the optimal Q-function $Q^*$ and the approximate Q-function $\hat{Q}$, respectively:

\begin{equation}
\begin{aligned}
Q^* &= \mathcal{T}Q^* \\
\hat{Q} &= \hat{\mathcal{T}}\hat{Q}
\end{aligned}
\end{equation}

We want to bound $\left|\hat{Q}(s,a) - Q^*(s,a)\right|$. Consider:

\begin{equation}
\begin{aligned}
&\left|\hat{Q}(s,a) - Q^*(s,a)\right| = \left|\hat{\mathcal{T}}\hat{Q}(s,a) - \mathcal{T}Q^*(s,a)\right| \\
&\leq \left|\hat{\mathcal{T}}\hat{Q}(s,a) - \hat{\mathcal{T}}Q^*(s,a)\right| + \left|\hat{\mathcal{T}}Q^*(s,a) - \mathcal{T}Q^*(s,a)\right|
\end{aligned}
\end{equation}

The first term can be bounded using the contraction property of the Bellman operator:

\begin{equation}
\begin{aligned}
\left|\hat{\mathcal{T}}\hat{Q}(s,a) - \hat{\mathcal{T}}Q^*(s,a)\right| &\leq \gamma \left|\hat{Q} - Q^*\right|
\end{aligned}
\end{equation}

For the second term, we have:
\begin{equation}
\begin{aligned}
&\left|\hat{\mathcal{T}}Q^*(s,a) - \mathcal{T}Q^*(s,a)\right| \\
&= \gamma \left| \mathbb{E}_{\hat{s}' \sim \hat{P}(\cdot|s,a)}\left[\max_{a'} Q^*(\hat{s}',a')\right] - \mathbb{E}_{s' \sim P(\cdot|s,a)}\left[\max_{a'} Q^*(s',a')\right] \right|
\end{aligned}
\end{equation}

Since $Q^*$ is Lipschitz continuous with constant $L_Q = \frac{L_r}{1-\gamma}$ (a standard result in RL), and the ASA error is bounded by $\epsilon$, we have:
\begin{equation}
\begin{aligned}
\left|\hat{\mathcal{T}}Q^*(s,a) - \mathcal{T}Q^*(s,a)\right| &\leq \gamma L_Q \epsilon = \gamma \frac{L_r}{1-\gamma} \epsilon
\end{aligned}
\end{equation}

Combining these results:
\begin{equation}
\begin{aligned}
\left|\hat{Q}(s,a) - Q^*(s,a)\right| &\leq \gamma \left|\hat{Q} - Q^*\right| + \gamma \frac{L_r}{1-\gamma} \epsilon
\end{aligned}
\end{equation}

Taking the supremum over all $(s,a)$:
\begin{equation}
\begin{aligned}
\left|\hat{Q} - Q^*\right| &\leq \gamma \left|\hat{Q} - Q^*\right| + \gamma \frac{L_r}{1-\gamma} \epsilon \\
(1-\gamma)\left|\hat{Q} - Q^*\right| &\leq \gamma \frac{L_r}{1-\gamma} \epsilon \\
\left|\hat{Q} - Q^*\right| &\leq \frac{L_r \epsilon}{(1-\gamma)^2}
\end{aligned}
\end{equation}

This completes the proof.
\end{proof}

\section{Experimental Settings and Algorithm Pseudocode}
\subsection{Evaluation Metrics}
\label{appdx:metrics}
The detailed evaluation metrics of compared methods are shown as follows:
\begin{itemize}
    \item \textbf{Session Length}: The number of requests in one session of a user with RS, which directly reflects the user satisfaction of the platform.
    \item \textbf{Watch Time}: The accumulated watching time of all items watched by a user in one session.
    \item \textbf{CTR}: The average click rate of all items recommended to a user in one session.
    \item \textbf{Diversity}: Quantifies the variety of content types in recommendations and is highly related to author benefits.
    \item \textbf{New Fans}: The total number of new followers accumulated by authors.
    \item \textbf{UA Count}: The count of user-author pairs with "deep" relationship which is defined as whether user has followed author, whether user has given author the most gifts, and other conditions.
    \item \textbf{Weekly Gifted Users}: The number of gifted user in a week and it indicates the revenue scale of users in the platform.
    \item \textbf{Total Revenue}: The important and ultimate metric to evaluate the platform revenue profits.
    \item \textbf{App Usage Time}: Average time users spend on the app.
    \item \textbf{Weekly Retention}: The stickness of user in a week. 
\end{itemize}

\subsection{Implementation Details}
\label{appendix:impl}
Notably, 
all the models adopt the same hyperparameters listed in Table \ref{tab:hyperparameters} for fair comparison.
\begin{table}[htbp]
  \centering
  \caption{Hyperparameters of the compared models.}
  \label{tab:hyperparameters}
  \fontsize{8}{11}\selectfont  
  \begin{tabular}{ll}
    \hline
    \textbf{Hyper-parameter} & \textbf{Value}  \\
    \hline
    Optimizer & Adam  \\
    $\gamma$ Discount factor & 0.9  \\
    $\tau$ Target network update rate & 0.005 \\
    Learning rate of critic & 1e-3  \\
    Learning rate of actor & 1e-4  \\
    Batch size & 1024 \\
    Train epochs & 250  \\
    Hidden layer dimensions & [64, 64]  \\
    The dimension of embedding layer & 32  \\
    Learning rate of embedding layer & 1e-3 \\
    Training steps per epoch  & 1e4 \\
    Training Platform & PyTorch \\
    \hline
    \end{tabular}
\end{table}

\subsection{Algorithm Pseudocode}\label{appdx:algo}
As shown in Algorithm \ref{algo1}, the training process of the RLIV-UA is divided into two steps: The pre-training of the state predictor $f_\theta$ and the joint training of the LIV networks and the tate predictor.

 \begin{algorithm}
    \caption{The RLIV-UA Algorithm} \label{algo1} 
    \begin{algorithmic}[1] 
        \Input{Initial parameters of all the Q networks $\phi_i,i=1,...,n$, initial parameters of state predictor $\theta$, learning rate $\lambda$, batch size for updating $B$ and the update rate of target networks $\tau$}
        \State \textbf{Initialize} the target networks with $\phi'_i \leftarrow \phi_i$
        \State \textbf{Initialize} the replay memory buffer of Q networks $D \leftarrow \emptyset$
        \State \textbf{Repeat:}
        \State  \textbf{Step1: The pre-training of state predictor}
        \For{$\text{epoch} = 1$ to \text{train\_epochs}}
            \For{each \text{interact\_step} $t$ of each user-author pair $\mathrm{ua}$}
            \State Observe the current state $\mathbf{s}^t_{\mathrm{ua}}$.
            \State Retrieve the last interaction from user historical behaviors sequences $\mathbf{s}^{t-1}_{\mathrm{ua}},a^{t-1}_{\mathrm{ua}},r^{t-1}_{\mathrm{ua}}$.
            \State Conduct gradient update for $f_\theta$ :$\theta\leftarrow\theta- \lambda \nabla_{\theta}\mathcal{L}_{\theta}(\theta)$
        \EndFor
        \EndFor
        \State \textbf{Step2: The join-training of LIV networks}
        \For{$\text{epoch} = 1$ to \text{train\_epochs}} 
            \For{each \text{interact\_step} $t$ of each user-author pair $\mathrm{ua}$}
                \State Observe the current state $\mathbf{s}^t_{\mathrm{ua}}$, action $a^t_\mathrm{ua}$ and rewards $r^t_{\mathrm{ua},i},i=1,...,n$.
                \State Input them into the state predictor $f_\theta$ to predict the next state $\hat{\mathbf{s}}^{t+1}_{\mathrm{ua}}$. 
                \Let{$D$}{$D \cup (\mathbf{s}^t_{\mathrm{ua}},a^t_\mathrm{ua},r^t_{\mathrm{ua}},\hat{\mathbf{s}}^{t+1}_{\mathrm{ua}})$}
                \State Retrieve the last interaction from user historical behaviors sequences $\mathbf{s}^{t-1}_{\mathrm{ua}},a^{t-1}_{\mathrm{ua}},r^{t-1}_{\mathrm{ua}}$.
            \State Conduct gradient update for $f_\theta$ :$\theta\leftarrow\theta- \lambda \nabla_{\theta}\mathcal{L}_{\theta}(\theta)$
                \If{the number of samples in $D$ reaches $B$}
                    \State sample a batch from $D$ and calculate total loss $\mathcal{L}$ shown in \ref{total_loss_function}
                    \State $\phi_i\leftarrow\phi_i-\lambda \nabla_{\phi_i}\mathcal{L}(\phi_i)$
                    \State $\phi'_i\leftarrow(1-\tau)\phi_i+\tau\phi'_i$
                \EndIf
            \EndFor
        \EndFor 
        \State \textbf{Return} The final Q networks parameters 
    \end{algorithmic}
\end{algorithm}

\section{Data Analysis} \label{appendixB}
 Firstly, we observe that the lifelong UA interactions have strong connection with the ultimate platform revenue. As shown in Fig. \ref{fig:sub1}, the count of UA pairs with both follow and frequent gift ("deep") relationship 
 is positively correlated with the total revenue value. 
On the other hand, as users continue to interact with the authors with "deep" relationship, its conversion rate will increase, as shown in Fig. \ref{fig:sub2}. 
 Hence, it provides insights to improve the platform profits by modeling the LIV of UA pairs.

 \begin{figure}[htbp]
  \centering
  \begin{subfigure}[t]{0.23\textwidth}
    \centering
    \includegraphics[width=\textwidth]{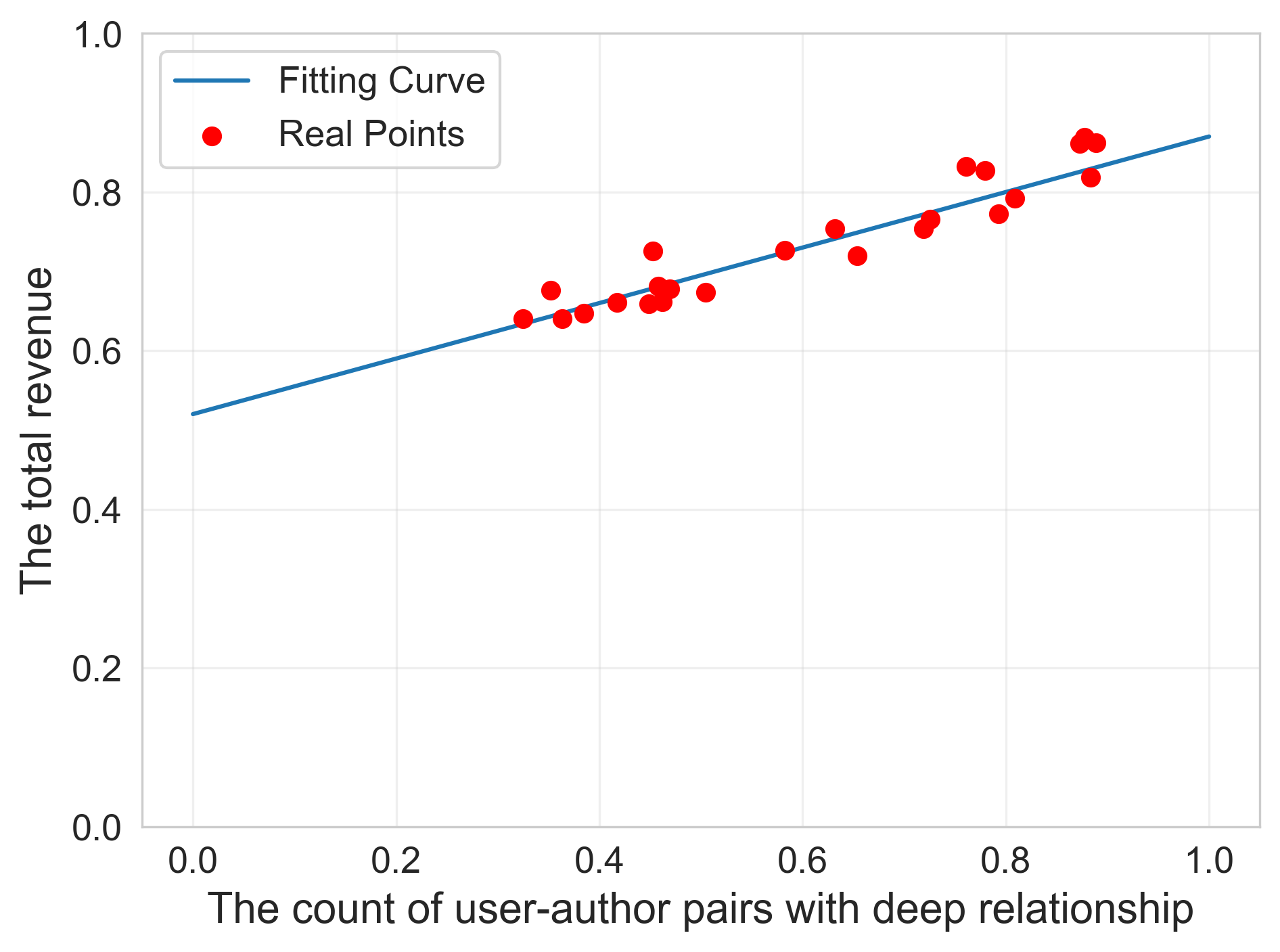}
    \caption{The relationship between the count of UA pairs with "deep" relationship and the total revenue.}
    \label{fig:sub1}
  \end{subfigure}
  \hfill
  \begin{subfigure}[t]{0.23\textwidth}
    \centering
    \includegraphics[width=\textwidth]{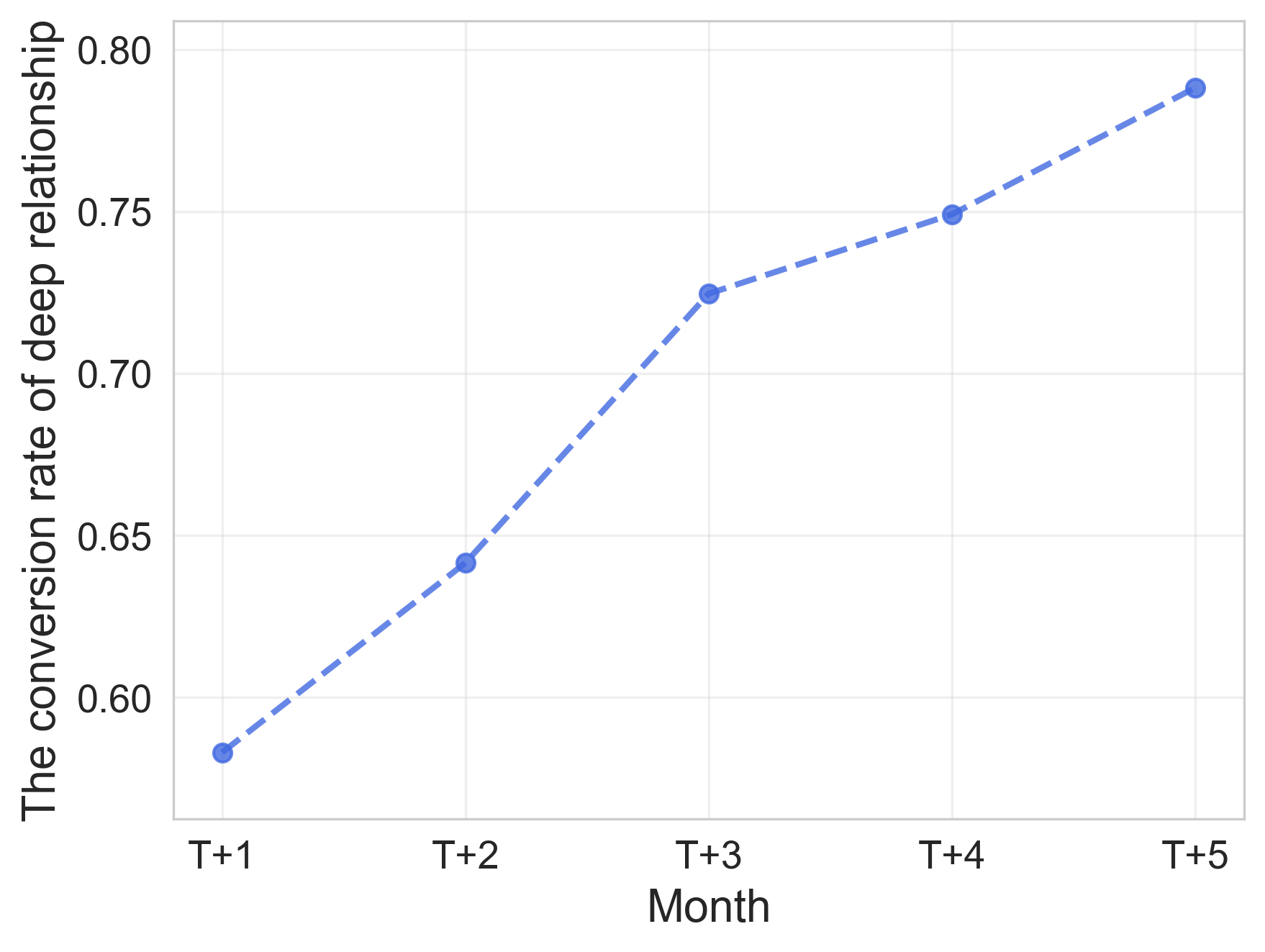}
    \caption{The conversion rate of UA pairs with "deep" relationship.}
    \label{fig:sub2}
  \end{subfigure}
  \caption{The revenue relevance and conversion rate of the lifelong UA relationship in Kwai app. Note that all data is collected in the second half of 2024 from Kwai app and scaled between 0 and 1.}
  \label{image0}
\end{figure}
 


\begin{figure}[htbp]
  \centering
  \includegraphics[width=0.65\linewidth]{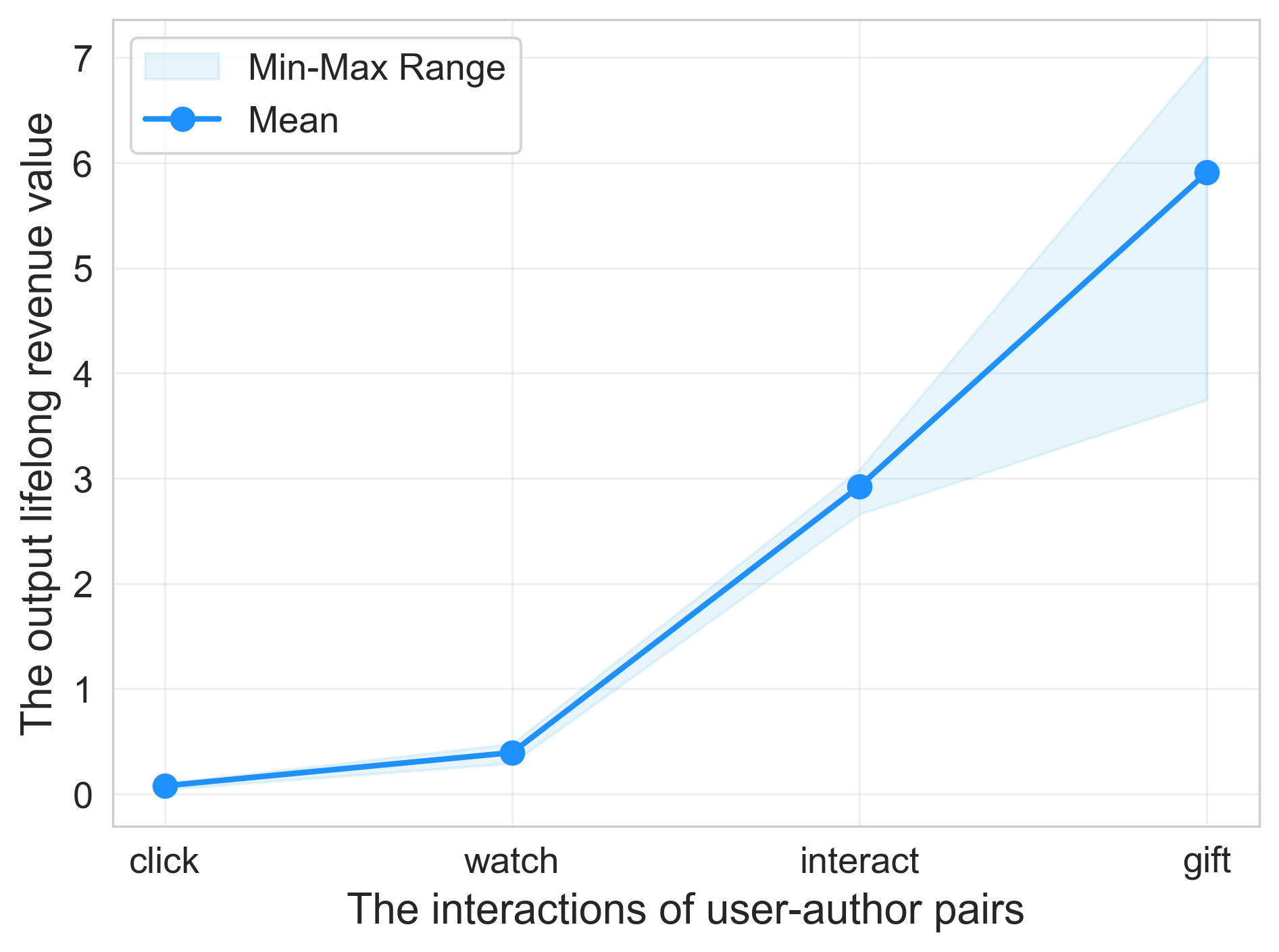}
  \caption{The output lifelong revenue values of some user-author pairs as their relationship becomes deeper.}
  \label{image6}
\end{figure}

In particular, to more clearly evaluate how the RLIV-UA model optimizes the LIV of UA pairs 
a specific example is shown in Fig. \ref{image6}. We collect the lifelong revenue value (Q value) output by the RLIV-UA model with progressive UA interactions from click, watch to interact, and gift.
It shows that as the UA output lifelong revenue value of user-author pair is progressively increasing as their relationship becomes deeper.
\section{More Discussion of ASA}
\label{appdx:asa}
\subsection{The rule-based approximation}
\label{appdx:rule}

The details of the rule-based approximation are as follows.

Specifically, the dynamic features $\mathbf{f}_{\mathrm{ua}}^t$ in the current state $\mathbf{s}^{t}_{\mathrm{ua}}$ is defined as the counts of several kinds of interactions between $\mathrm{ua}$. Note that $\mathbf{f}_{\mathrm{ua}}^{t + 1}$ can be approximated based on user's current rewards $r^t_{\mathrm{ua}}$ and current dynamic UA features $\mathbf{f}_{\mathrm{ua}}^{t}$: 
\begin{equation}
\mathbf{f}^{t+1}_{ua} = 
\begin{cases}
\mathbf{f}_{ua}^t+1, & \text{if } r^t_{ua}>0 \\
\mathbf{f}_{ua}^t,  & \text{otherwise }
\end{cases}
\end{equation}
Since the SCRI-MDP only focuses on the interactions and state transitions between UA pair like $\mathrm{ua}$, it ignores interactions between user $u$ and items of other authors. Under the above assumption of the SCRI-MDP, the next state $\hat{\mathbf{s}}^{t+1}_{ua}$ can be derived by the dynamic features $\mathbf{f}^{t+1}_{ua}$.

\subsection{Offline evaluation of ASA}
\begin{table}[htbp]
  \centering
  \caption{The evaluation loss of the adjacent state approximation compared with rule-based approximation.} \label{asa}
  \label{tab:onlinetime}
  \fontsize{8}{11}\selectfont
  \begin{tabular}{lll}
    \hline
    \textbf{Methods} & 
    \textbf{MSE loss} & 
    \textbf{Cosine Similarity loss} \\
    \hline
    rule-based approximation & 0.718 & 0.743   \\
    proposed ASA method & \textbf{0.502} & \textbf{0.630} \\
    \hline
    \end{tabular}
\end{table}
To demonstrate the effectiveness of the ASA method, we conduct a comparative evaluation against a rule-based baseline on an offline test set. The full dataset is randomly split into training and test sets in an 8:2 ratio. As shown in Table \ref{asa}, the ASA method achieves significantly lower prediction error across all key state features, confirming its superior ability to model the complex, long-term dynamics of user-author interactions compared to the heuristic rule-based approach.
\end{document}